\documentclass[prb,twoside,twocolumn,showpacs]{revtex4}
\usepackage{graphicx}
\usepackage{amsmath}
\usepackage{amssymb}

\begin{document}

\title{Distinct order of Gd $4f$ and Fe $3d$ moments coexisting in GdFe$_4$Al$_8$}
\author{M. Angst}
 \email[Email: ]{angst@ameslab.gov}
\author{A. Kreyssig}
\author{Y. Janssen}
\author{J.-W. Kim}
\author{L. Tan}
\author{D. Wermeille}
\author{Y. Mozharivskyij}
\author{A. Kracher}
\author{A.~I. Goldman}
\author{P.~C. Canfield}
%\author{}
%\author{}
\affiliation{Ames Laboratory DOE and Department of Physics and
Astronomy, Iowa State University,Ames,IA 50011,USA}

\date{\today}
\begin{abstract}
Single crystals of flux-grown tetragonal GdFe$_4$Al$_8$ were
characterized by thermodynamic, transport, and x-ray resonant
magnetic scattering measurements. In addition to antiferromagnetic
order at $T_N \approx 155\, {\rm K}$, two low-temperature
transitions at $T_1 \approx 21\, {\rm K}$ and $T_2 \approx 27\,
{\rm K}$ were identified. The Fe moments order at $T_N$ with an
incommensurate propagation vector ($\tau , \tau, 0$) with $\tau$
varying between $0.06$ and $0.14$ as a function of temperature,
and maintain this order over the entire $T<T_N$ range. The Gd $4f$
moments order below $T_2$ with a ferromagnetic component mainly
out of plane. Below $T_1$, the ferromagnetic components are
confined to the crystallographic plane. Remarkably, at low
temperatures the Fe moments maintain the same modulation as at
high temperatures, but the Gd $4f$ moments apparently do not
follow this modulation. The magnetic phase diagrams for fields
applied in [110] and [001] direction are presented and possible
magnetic structures are discussed.
\end{abstract}


\maketitle

\section{Introduction}
\label{intro}

\begin{figure}[tb]
\includegraphics[width=0.98\linewidth]{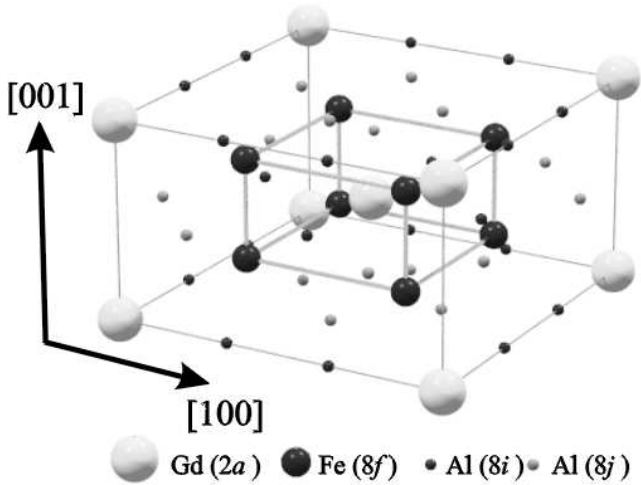}
\caption{(Color online) The crystal structure of $R$Fe$_4$Al$_8$.
Thin lines outline the tetragonal (I$4/$mmm) unit cell, thick
lines denote the Fe ``cage'' around the rare earth $R$. }
\label{Fig1}
\end{figure}

Understanding the interplay of rare-earth local-moment magnetism
and $3d$ transition metal itinerant magnetism is of fundamental
physical interest\cite{Campbell72} and may help in the design of
more efficient permanent magnets.
The tetragonal $R$Fe$_4$Al$_8$ ($R=$rare earth) compounds %sdf
(Fig.\ \ref{Fig1}) are well suited for studying this interplay
because of simple symmetry conditions and because the interaction
between the two magnetic sublattices is rather weak: Fe moments
appear to order without a corresponding $R$ $4f$ order.
Consequently, numerous studies have been performed on
$R$Fe$_4$Al$_8$ (R148) over the last 30 years and interest in
these compounds has remained
high.\cite{Buschow78,Felner78,Fujiwara87,Duong01,Duong02,Langridge99,Paixao00,Paixao01,Schobinger98,Schobinger99}

Neutron scattering studies performed on R148 with various $R$
indicate that Fe moments, between which the coupling is strongest,
order between $100$ and $200\, {\rm K}$, generally in cycloidal
structures with propagation vectors parallel to [110] and moments
confined to the $a$$-$$b$
plane.\cite{Paixao00,Paixao01,Schobinger98,Schobinger99} For
magnetic rare earths $R$, ordering of the $R$ moments, with much
weaker coupling and at much lower temperature $T$, has been
reported. However, the critical temperature associated with this
ordering is very poorly defined. For example in Dy148 at
$\sim$$50\,{\rm K}$ the magnetic dc susceptibility rises faster
than expected for Curie-Weiss behavior, but there are no sharp
features.\cite{Paixao00} Furthermore, the rare earth moments were
found to follow the Fe-moment
modulation.\cite{Paixao00,Schobinger98} This seems to imply that
the Fe$-R$ moment interaction is stronger than the $R$$-$$R$
moment interaction.

Of all rare earth elements, Gd has two highly interesting specific
features associated with its high-spin and zero-angular-momentum
$4f$ state: 1) high spin results in strong magnetic coupling
between the localized $4f$ and the conduction electrons, implying
large magnetic interactions and the largest de Gennes factor of
all rare earths; 2) zero angular momentum implies a spherical $4f$
charge-cloud and no magneto-crystalline anisotropy (MCA) resulting
from crystal-electric-field effects $-$ the direct interplay of
magnetic interactions can be studied without
crystal-electric-field effects. In view of this, it is somewhat
surprising that GdFe$_4$Al$_8$ (Gd148) has been much less
studied\cite{Fujiwara87,Duong01} than R148 with other rare earths.
In particular, apart from M{\"{o}}ssbauer
spectroscopy,\cite{Buschow78} there have been no microscopic
studies of Gd148.

Here, we report on the flux-growth of single crystals of the
GdFe$_4$Al$_8$ phase, and on the characterization of the crystals
by magnetization, electrical transport, specific heat, and x-ray
diffraction. We also report on first synchrotron x-ray resonant
magnetic scattering (XRMS) data. We provide evidence for two
consecutive phase transitions at low temperature $T$ in addition
to the N\'{e}el transition at $155\, {\rm K}$, and show that the
two, low-$T$, transitions are associated with the ordering of the
Gd $4f$ moments, resulting in complex magnetic structures
involving both ferro- and antiferromagnetic components. In
contrast to other R148 compounds, the $4f$-moment ordering is
associated with a propagation vector that is distinct from that
associated with the Fe moment order, which at low $T$ has the same
modulation as in the vicinity of the {N{\'{e}}el} temperature. We
also present the magnetic phase diagram for two field directions
and discuss the implications of changes in the Fe stoichiometry.

The paper is organized into seven sections. In Sec.\ \ref{exp} we
describe the flux growth of the single crystalline samples used in
the study, and the experimental procedures. In Sec.\ \ref{lowH}
zero and low field electrical transport, magnetization, and
specific-heat data are presented, and we evaluate the influence of
the iron stoichiometry on the physical properties. In Sec.\
\ref{highH} we present electrical transport and magnetization data
measured in fields for two field directions, and in Sec.\
\ref{xrms} the results of a first XRMS experiment are described.
Finally, we discuss the low-temperature magnetic phase diagram and
possible magnetic structures in Sec.\ \ref{disc}, before
summarizing our main conclusions in Sec.\ \ref{conc}.

\section{Experimental}
\label{exp}

Whereas all previous single crystals of $R$Fe$_4$Al$_8$ material
were obtained using the Czochralsky-method, we have grown single
crystals of Gd148 (and other R148) with a self-flux
method.\cite{Sol1,Sol2,Sol3} One of the problems with studies on
single crystals of R148 is a width of formation often observed,
involving some Fe atoms occupying nominal Al positions and vice
versa; this can lead to striking differences in the magnetic
structure and phase diagram as compared to the stoichiometric
material (see, e.g., Refs.\ \onlinecite{Paixao01,Waerenborgh00}).
A flux-growth procedure may allow better control of the
stoichiometry of the crystals by varying the starting composition.
Unfortunately, information about the ternary phase diagram
$R$-Fe-Al is very limited. We used differential thermal analysis
(DTA; see Ref.\ \onlinecite{Janssen05} for a review of the use of
DTA in the flux growth of crystals) to establish i) that
GdFe$_4$Al$_8$ is congruently melting and ii) selected solidus
temperatures in the ternary around the GdFe$_4$Al$_8$
stoichiometry. For the crystals used in this study, we selected
starting compositions between Gd148 and the also congruently
melting Fe$_2$Al$_5$ (Gd$_3$Fe$_{16}$Al$_{34}$ and
Gd$_2$Fe$_{14}$Al$_{31}$), which should not lead to large
imbalances in the Fe to Al ratio.

Elements in ratios corresponding to the starting composition were
first arc-melted together. The resulting ingot was placed in an
alumina crucible, which was wrapped in Ta foil (to prevent any
residual oxygen in the Ar from reacting with the sample) and
placed in a vertical tube furnace in flowing Ar. Crystals were
grown by heating to $1475^{\circ}{\rm C}$ and then slowly
($2^{\circ}{\rm C} / {\rm h}$) cooling to $1180^{\circ}{\rm C}$,
at which point the furnace was turned off. The flux was removed
from the crystals in a second step by heating to $1200^{\circ}{\rm
C}$, keeping the temperature for $30\,{\rm min}$ and then
decanting, following procedures described in Refs.\
\onlinecite{Sol1,Sol2,Sol3,Janssen05}.

For both starting compositions, we obtained crystals of the Gd148
phase, as identified via powder x-ray diffraction. Crystals
typically grow prismatically with (110) facets and the long
direction parallel to [001], as determined with Laue scattering.
We obtained crystals of dimensions up to about $10\times 2\times 2
{\rm mm}^3$. We determined, via single crystal
x-ray-diffraction-structure refinement and electron-microprobe
analysis (employing single crystals of GdFe$_2$ as standards), the
crystals to be slightly iron deficient, but detected no Al on Fe
sites or vice versa (measured compositions were between
GdFe$_{3.88(5)}$Al$_8$ and GdFe$_{3.96(1)}$Al$_8$; deviations of
the Al stoichiometry from $8$ were always less than 2 standard
deviations and are not listed).

We note that the x-ray-diffraction-structure refinement cannot
distinguish between a Fe deficiency due to vacancies and a
(larger) Fe deficiency due to Al on Fe sites, both scenarios can
give an equal electron density. However, assuming an Al/Fe mixture
to retain the full occupancy of the Fe site leads (for one of the
investigated crystals) to a GdFe$_{3.92(1)}$Al$_{8.08(1)}$
composition and, thus, to an excess of Al. Since the
electron-microprobe analysis results indicated no significant
variations in the $1/8$ ratio of Gd to Al, we did the structure
refinements under the assumption that the electron density on the
Fe sites is caused by vacancies, although some Al/Fe substitution
could never be excluded. Whatever scenario is chosen, none of our
conclusions drawn in this article are affected by the issue of
whether or not some Al atoms are present on the Fe site, and for
definiteness subsequently we will assume that the vacancy scenario
is the correct one. The composition of our samples can thus be
described with the empirical formula GdFe$_{4-\delta}$Al$_8$, with
$\delta \approx 0.04-0.12$. Surface scans in the
electron-microprobe analysis gave no indications of compositional
variations within the same crystal.

Since the exact iron stoichiometry varies slightly from crystal to
crystal (even within the same growth batch) and this was found to
have a significant influence on the magnetic properties (see Sec.\
\ref{lowH} below) most of the measurements presented below were
performed on the same crystal, sample I, which has a refined
composition GdFe$_{3.96(1)}$Al$_8$ ($a=8.7699(9)\, $\AA,
$c=5.0440(6)\, $\AA). A bar cut from the crystal by a wire saw was
connected with contacts for 4-point electrical-transport
measurements with the current density $j\|$[110] (sample Ia).
Other samples were connected with contacts for electrical
transport measurements with either $j\|$[110] or $j\|$[001]. On a
second bar (sample Ib, which had a mass of $2.22\,{\rm mg}$) cut
from crystal I we measured the longitudinal magnetization and the
zero-field specific heat. In order to investigate the relation
between compositional variations and variations in the transition
temperatures, we carried out additional magnetization measurements
on several other crystals, the composition of which was determined
by x-ray diffraction and electron microprobe. The additional
crystals are labelled II to VII in the order of their appearance
in the text. The above measurements were performed with commercial
(Quantum Design) laboratory equipment. For the high-field
measurements, the sample orientation was fixed with a
two-component glue (X60 from {W{\"{a}}getechnik} GmbH, Darmstadt,
Germany).

Sample Ib was also used in the XRMS experiment, performed on the
6ID-B undulator beamline in the MUCAT sector at the Advanced
Photon Source, Argonne National Laboratory. The incident energy of
x-rays was tuned to the Gd $L_{\rm II}$ edge ($E=7.934\,{\rm
keV}$), using a liquid nitrogen cooled, double crystal Si (111)
monochromator and a bent mirror. Sample Ib was mounted on a copper
rod on the cold finger of a closed cycle displex refrigerator,
such that a natural (110) facet was exposed to the x-ray beam.
Thermal transfer was enhanced by embedding the sample in copper
paste. The sample, oriented so that the scattering plane of the
experiment was coincident with the $a-b$ plane, was encapsulated
in a Be dome with He exchange gas to further enhance thermal
transfer. The incident radiation was linearly polarized
perpendicular to the scattering plane ($\sigma$ polarized). In
this geometry, only the component of the magnetic moment that is
in the scattering plane contributes to the resonant magnetic
scattering arising from electric dipole transitions ($E1$) from
$2p$ to $5d$ states. Furthermore, the dipole resonant magnetic
scattering rotates the linear polarization into the scattering
plane ($\pi$ polarization). In contrast, charge scattering does
not change the polarization of the incident beam
($\sigma\rightarrow\sigma$ scattering). Pyrolytic graphite PG
(006) was used as a polarization analyzer, selecting primarily
$\pi$ polarized radiation. For $E=7.934\,{\rm keV}$, the
polarization analyzer used reduces the detected intensity
resulting from $\sigma\rightarrow\sigma$ charge scattering by
about $99.9\%$, whereas the $\sigma\rightarrow\pi$ resonant
magnetic scattering is passed with little loss. Thus, the
polarization analysis suppresses the charge scattering relative to
the magnetic scattering signal.

\label{res}
\section{Low field measurements and influence of Fe content}
\label{lowH}
\begin{figure}[tb]
\includegraphics[width=0.98\linewidth]{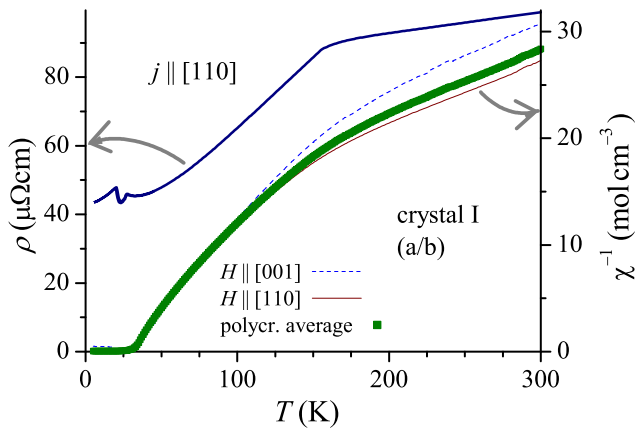}
\caption{(Color online) Zero-field resistivity $\rho$ (along
[110], sample Ia) and inverse dc susceptibility $\chi ^{-1}$ vs
temperature $T$ (sample Ib). } \label{Fig2}
\end{figure}

In this section, we will first present zero-field transport and
low field magnetization measurements on our main crystal (samples
Ia/Ib). Then, we will briefly discuss sample-to-sample differences
in the characteristic temperatures, which we think are related to
Fe deficiency. Finally, we will present specificheat measurements
on sample Ib indicating that the Gd $4f$ moments are ordered only
at low temperatures. The main implications of the measurements
will be discussed later in Sec.\ \ref{disc}.

In Fig.\ \ref{Fig2}, we show, for sample Ia, the $T$ dependence of
the resistivity $\rho$ ($j\|$[110]). Since the sample is rather
small, the estimated uncertainty in the absolute value of the
resistivity is about 15\%. The antiferromagnetic ordering at $T_N
\simeq 155\, {\rm K}$ is visible as a sharp kink in $\rho
(T)$.\cite{note_Palasyuk04} The inverse dc magnetic susceptibility
$\chi^{-1}$ ($M/H$) of sample Ib, cut from the same crystal as
sample Ia, is also presented in Fig.\ \ref{Fig2}. The dashed and
dotted curves shown are determined from the magnetization $M(T)$
in a field of $\mu_0 \,H=0.1\,{\rm T}$ applied parallel to [110]
and parallel to [001]. Below this field $M$ vs $H$ is linear for
$T \gtrsim 40\, {\rm K}$. Since $\chi$ is slightly anisotropic, we
also calculated the polycrystalline average $\chi_{\rm
poly}=(2\chi_{110}+\chi_{001})/3$, shown as squares. In
$\chi^{-1}$ for both directions the antiferromagnetic ordering is
manifest by a change in slope, but it is much less sharp than the
corresponding signature in the resistivity. A weak signature of
the {N{\'{e}}el} transition may be expected if, as we will discuss
below, only the Fe moments order at $T_N$. In agreement with
measurements on polycrystalline samples,\cite{Buschow78} $\chi$
follows Curie-Weiss behavior both above and, over a limited $T$
range, below $T_N$. We found (on the calculated polycrystalline
average) above $T_N$ a Weiss temperature of $-165\,{\rm K}$ and an
effective moment of $11.4 \, \mu_B \, / {\rm f.u.}$. Assuming
contributions by the magnetic atoms that are additive in the Curie
constant and a contribution by Gd of $7.9 \, \mu_B \, / {\rm Gd}$,
the free moment per Fe atom in the paramagnetic state is $4.1 \,
\mu_B \, / {\rm Fe}$. Below $T_N$ (range $70-115\,{\rm K}$) we
found a Weiss temperature of $17\,{\rm K}$ and an effective moment
of $7.4 \, \mu_B \, / {\rm f.u.}$. The latter is closer to the Gd
free-ion value ($7.9 \, \mu_B \, / {\rm f.u.}$) than the one
reported by Buschow {\em et al.}\cite{Buschow78} ($6.2 \, \mu_B \,
/ {\rm f.u.}$).

\begin{figure}[t!b]
\includegraphics[width=0.98\linewidth]{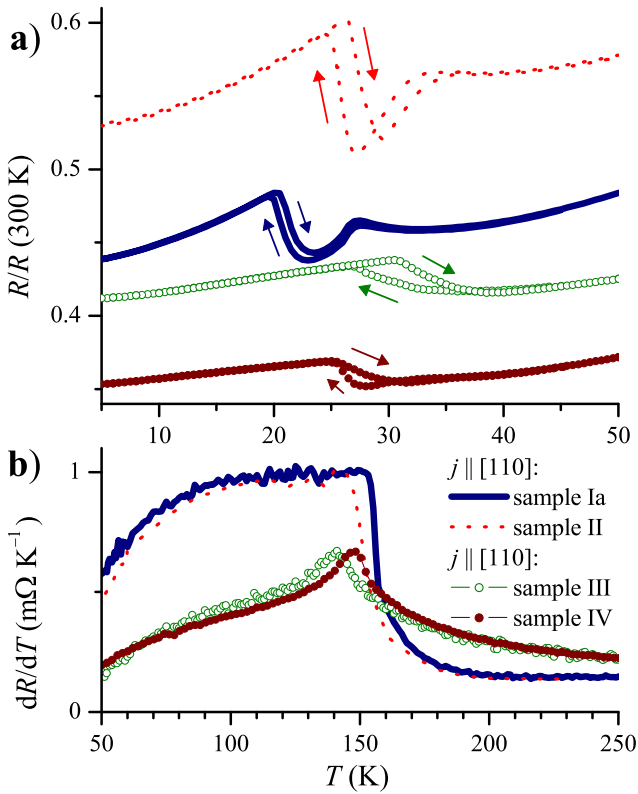}
\caption{(Color online) Low-temperature normalized resistance
(panel a) and derivative of high temperature resistance (panel b)
vs temperature $T$ of various crystals: Ia (thick line) and II
(dotted line) for $j\|$[110], III (open circles) and IV (full
circles) for $j\|$[001].} \label{Fig3}
\end{figure}

The resistivity curve shows additional features just below $30\,
{\rm K}$, magnified in Fig.\ \ref{Fig3}a) (thick line): The broad
peak in $\rho (T)$ at $\sim \! \! 28\, {\rm K}$ with a drop at
$\sim \! \! 26.5\,{\rm K}$ ($T_2$) indicates a phase transition.
Furthermore, with decreasing $T$, at $\sim \! \! 21\, {\rm K}$
($T_1$) there is a sudden increase in $\rho$ by $\sim 10\%$. This
feature is hysteretic in temperature, suggesting that the feature
is associated with a first-order transition.\cite{note_Fujiwara}
Note that the hysteresis visible at $T_2$ is opposite to what is
expected for a first-order transition (i.e.,
superheating/supercooling). It is most likely a remnant of the
hysteresis associated with the $T_1$ transition. At $T_1$, the
form of the hysteresis is consistent with
superheating/supercooling.

Also shown in Fig.\ \ref{Fig3}a) are low-$T$ zero-field
resistivity curves for three additional samples, one also with
$j\|$[110] and two with $j\|$[001]. Whereas the feature of a
sudden increase of $\rho$ with decreasing $T$ is visible in all
curves, the other feature at slightly higher temperatures is only
identifiable when $j\|$[110]. Based on the results from sample Ia
only, it may be tempting to associate the increase in $\rho$ (with
decreasing $T$) with the opening of a superzone
gap.\cite{Mackintosh62} However, the presence of this increase
with the same order of magnitude for both $j\|$[110] and
$j\|$[001] makes a superzone gap scenario less likely.

It is also clear from Fig.\ \ref{Fig3}a) that there are
significant variations of the temperatures $T_1$ corresponding to
this feature. For comparison, Fig.\ \ref{Fig3}b) shows the
resistivity derivative ${\rm d} \rho / {\rm d} T$ for the same
samples at higher $T$, in order to make the {N{\'{e}}el}
transition more visible, which manifests itself differently for
the two current density directions. The curves indicate a sharp,
well defined $T_N$ for each sample and comparing Figs.\
\ref{Fig3}a) and \ref{Fig3}b) suggests a correlation between
higher $T_1$ and lower $T_N$.

\begin{figure}[tb]
\includegraphics[width=0.98\linewidth]{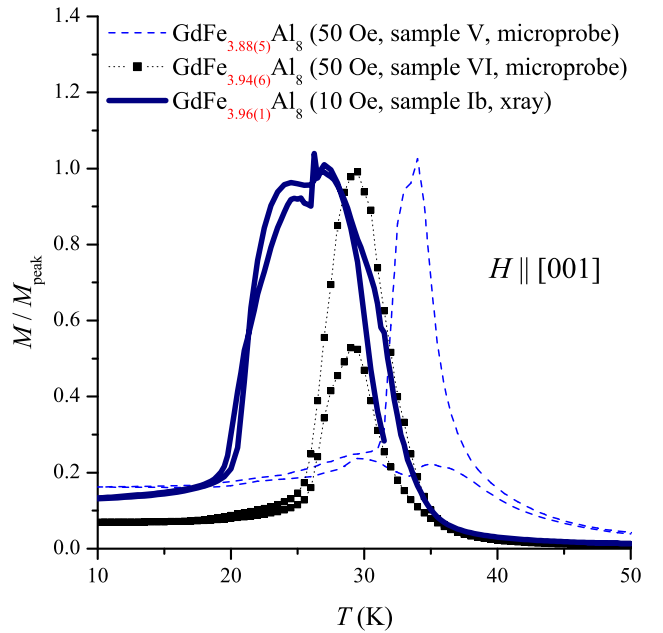}
\caption{(Color online) Low field ($H\|[001]$) Magnetization of
various crystals with Fe stoichiometry (as described in Sec.\
\ref{exp}) determined by electron microprobe or
x-ray-diffraction-structure refinement.} \label{Fig4}
\end{figure}

Figure \ref{Fig4} displays low field ($H\|$[001]) magnetization
measurements performed on three samples with different Fe
stoichiometry as determined by x-ray diffraction and electron
microprobe: samples Ib (GdFe$_{3.96(1)}$Al$_8$, thick line), V
(GdFe$_{3.94(6)}$Al$_8$, full squares), and VI
(GdFe$_{3.88(5)}$Al$_8$, dashed line). The data were taken for
both decreasing and increasing $T$ (field cooled), and
normalized\cite{note_normalizedM} to the maximum in magnetization.
A comparison with Fig.\ \ref{Fig3}a) indicates that the rise in
$M$ with $T$ is associated with $T_1$, the subsequent decrease
with $T_2$. A decrease in the Fe content seems to systematically
shift $T_1$ and $T_2$ to higher temperatures. We note that for
crystal I $M(T)$ (Fig.\ \ref{Fig4}) indicates a higher $T_2$ than
$\rho (T)$ [Fig.\ \ref{Fig3}a)]. In principle, this might be due
to a different Fe content of the two pieces (Ia and Ib) cut from
the crystal and used for these measurements. However, we do not
believe that this is the case, because the specific-heat peak
associated with $T_2$ (see below; measured on Ib) corresponds well
to $T_2$ deduced from $\rho (T)$ on sample Ia.

Combining the above data it seems that decreasing the Fe content
results in a decrease of $T_N$ and an increase of $T_1$ and $T_2$.
In order to confirm this, a study involving samples that are much
more Fe deficient, would be highly desirable.

\begin{figure}[tb]
\includegraphics[width=0.98\linewidth]{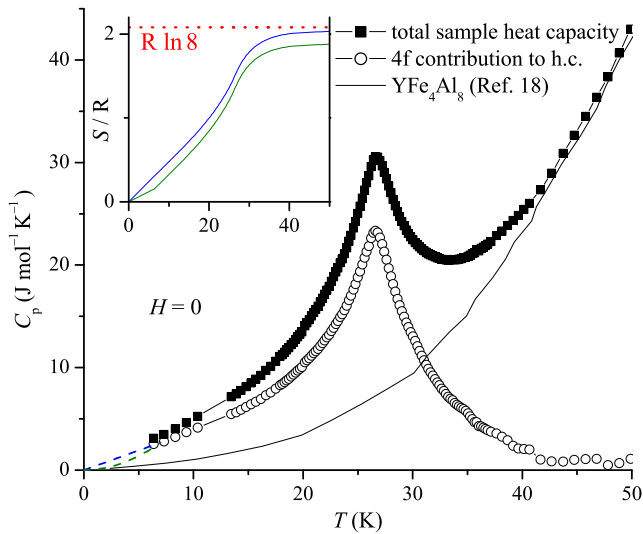}
\caption{(Color online) %
Specific heat $C_p$ measured on sample Ib in zero field
($\blacksquare$) and Gd $4f$ moment contribution to the specific
heat ($\circ$), obtained by subtracting the specific heat of
YFe$_4$Al$_8$ (full line, from Ref.\ \onlinecite{Hagmusa98}).
Inset: Entropy of the Gd $4f$ moments obtained by integrating the
$4f$ moment contribution to $C_p / T$. The two lines shown were
calculated using two extrapolations of the specific heat to $T=0$,
indicated by dashed lines in the main panel (see text for details
of the extrapolations). The dotted line in the inset indicates the
full Gd $4f$ entropy.} \label{Fig5}
\end{figure}

Figure \ref{Fig5} shows the measured specific heat $C_p$ of sample
Ib (closed squares) after subtraction of the addenda contribution
(the specific heat of sample platform and grease measured before
mounting the sample). Also shown is the mass-scaled specific heat
of YFe$_4$Al$_8$ from Ref.\ \onlinecite{Hagmusa98} (full line),
and the difference between the two specific heats (open circles).
Since the necessary mass-scaling is small, the remaining specific
heat is close to the magnetic contribution associated with the Gd
sublattice. The broad, asymmetric, peak at $\sim \! \! 26.5 \,
{\rm K}$ corresponds to the drop in $\rho$ at $T_2$ [determined on
sample Ia cut from the same crystal as sample Ib, see Fig.\
\ref{Fig3}a) dashed curve]. The broadness of the specific-heat
peak may suggest a degree of inhomogeneity of the Fe distribution
in the sample, although the electron-microprobe analysis provided
no indications of inhomogeneities within crystals.

No feature in the specific heat is visible around $T_1$, the small
discontinuity at $20\,{\rm K}$ indicated by the data was found to
be an artifact due to the change of the heat pulse intensity
applied by the system. We checked the specific-heat raw data, and
can exclude any latent heat restricted to a temperature region
smaller than the spacing of measurement points. We conclude that
the latent heat of the $T_1$ transition has to be small, i.e., the
phases above and below $T_1$ have similar entropies. The
alternative explanation of the specific heat, that $T_1$ is not a
first-order transition, seems unlikely in light of not only the
presence, but also the shape of thermal hysteresis observed in
$\rho (T)$, which is particularly clear when a magnetic field is
applied in the [110] direction (see Sec.\ \ref{highH} below).

The corresponding magnetic entropy $S$, obtained by integration,
is shown in the inset of Fig.\ \ref{Fig5}. The two curves shown
are calculated with two different extrapolations of $C_p$ to zero
temperature (dashed lines in the main panel). One extrapolation is
obtained by connecting the $C_p$ data point at the lowest $T$
linearly to $C_p=0$ at zero temperature, the other by connecting
the $C_p/T$ data at the lowest $T$ linearly to $C_p/T=0$ at zero
temperature. Given typically observed specific heats of Gd
compounds,\cite{note_DuongGdFe2Ge2} the actual specific heat at
low temperature will most likely be between these two
extrapolations, and thus the two curves shown in the inset of
Fig.\ \ref{Fig5} may be considered lower and upper limits of the
Gd $4f$ entropy (neglecting additional experimental
uncertainties). At $30\, {\rm K}$ the magnetic entropy already
reaches 80\% (78\% and 86\% for the two low-$T$ extrapolations
made) of the full entropy of Gd $4f$ moments (${\rm R}\,\ln 8$,
dashed line). Above $45\, {\rm K}$ it hardly varies anymore,
having reached $>90\%$ and $>97\%$ of the full $4f$ entropy for
the two low-$T$ extrapolations made. This strongly suggests that
the Gd $4f$ moments are not ordered above $T_2$, and the
transition at $T_2$ corresponds to the ordering of the $4f$
moments.

By careful measurement and analysis of the low-field thermodynamic
and transport data of solution grown single crystals of Gd148 we
have shown that there are sharp features associated with two
magnetic phase transitions. These transitions appear to be
primarily related to the ordering of the Gd sublattice. In
addition we have been able to establish a clear correlation
between small variations in the Fe stoichiometry and $T_N$, $T_1$,
and $T_2$. Having done this, we will now focus on the sample that
is closest to full stoichiometry (crystal I) and try to more fully
delineate and understand the field and temperature dependence and
the structure of the magnetic phases in this sample.

\section{High-field measurements}
\label{highH}

Applying a magnetic field often helps to clarify the nature of
magnetic phases observed in zero (or low) field measurements. We
therefore measured the field dependent electrical resistivity and
the magnetization on several samples. Shown in this section are
electrical transport (subsection \ref{highHrho}) and magnetization
(subsection \ref{highHM}) data measured on samples Ia/Ib at low
temperatures in fields applied in-plane and out-of-plane. The
features in resistivity corresponding to the two low-temperature
transitions are much sharper when an in-plane field is applied and
observed features suggest complex domain effects. The
magnetization data are used to estimate the in-plane and
out-of-plane components of the spontaneous ferromagnetic moment.
The phase diagrams for $H\|$[110] and $H\|$[001], which can be
constructed from these transport and magnetization data, will be
presented and discussed in Sec.\ \ref{disc}.

\subsection{Electrical transport}
\label{highHrho}
\begin{figure}[tb]
\includegraphics[width=0.98\linewidth]{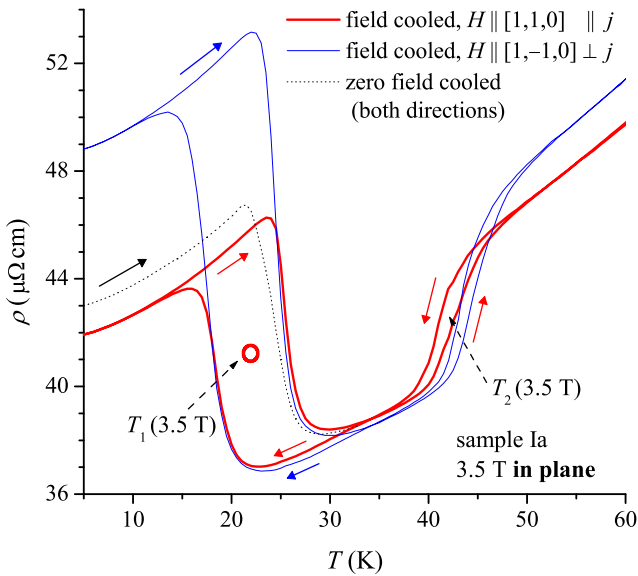}
\caption{(Color online) resistivity $\rho$ vs $T$ in fields of
$3.5\,{\rm T}$ applied in-plane parallel or perpendicular to the
current density $j\|[110]$ (sample Ia). Arrows indicate the $T$
direction in which the measurements were performed. The open
circle indicates the value of the $T_1$ transition in $3.5\,{\rm
T}$, averaged for $T$ increasing and decreasing.} \label{Fig6}
\end{figure}

Figure \ref{Fig6} shows the resistivity $\rho$ ($j\|$[110]) of
sample Ia when a field of $3.5\,{\rm T}$ is applied either $\|j$
(longitudinal resistivity) or $\|[1\overline{1}0]$ (i.e.\
transverse resistivity, $\perp$$j$, but in a crystallographically
equivalent in-plane direction). For both field directions, the
magnitude of the jump at $T_1$ is enhanced and the hysteresis in
temperature greatly increased, making the first-order nature of
the transition more apparent. The average
($T\uparrow$,$T\downarrow$) temperature of the $T_1$ feature,
however, is only weakly influenced by an in-plane field. In
contrast to this, the $T_2$ feature is shifted to higher $T$ by
$H\|$[110], suggesting a ferromagnetic nature of the phase at
$T<T_2$. Furthermore, the fields remove the peak in $\rho$ just
above the drop at $T_2$, and cause the feature to exhibit
hysteresis. This different response to $H\|$[110] confirms that
there are indeed two independent magnetic transitions $T_1$ and
$T_2$, dividing the ordered state below $T_N$ into three magnetic
phases, a low-temperature phase (LTP), an intermediate-temperature
phase (ITP), and a high-temperature phase (HTP).

\begin{figure}[tb]
\includegraphics[width=0.98\linewidth]{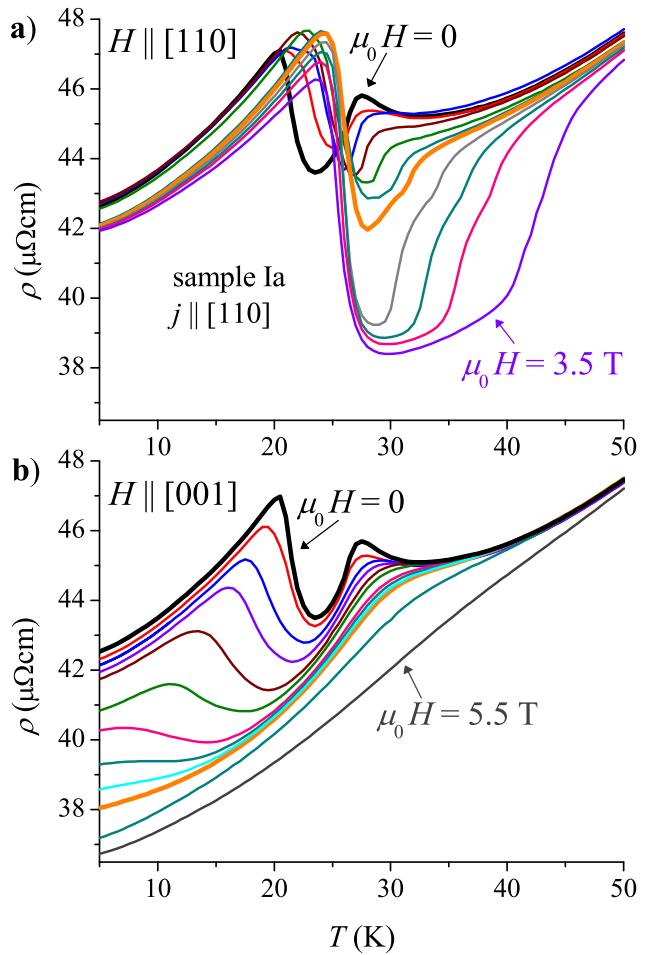}
\caption{(Color online) a) Resistivity $\rho$ vs $T$ (sample Ia,
$j\|$[110], $T$ increasing) in fields applied parallel to
$j\|$[110]. From top to bottom 0 (thick line), 0.1, 0.2, 0.4, 0.5,
0.8, 1 (thick line), 1.5, 2, 2.5, $3.5\,{\rm T}$. b) As panel a),
but in fields applied parallel to [001] ($j\|$[110]). From top to
bottom 0 (thick line), 0.1, 0.2, 0.3, 0.4, 0.5, 0.7, 0.8, 0.9, 1,
2, $5.5\,{\rm T}$.} \label{Fig7}
\end{figure}

Whereas there are similarities in $\rho (T)$ for the two field
directions considered, there are also very pronounced differences,
particularly for the $T_1$ feature, below which the field cooled
$\rho(T)$ is much higher for $H$$\perp$$j$ than for $H\|j$. A
transverse magnetoresistance higher than the longitudinal one
might arise from a bending of the electron trajectory by the
Lorentz force (see, e.g., Ref.\ \onlinecite{Harris95}). However,
the zero-field-cooled $\rho (T)$ below $33\,{\rm K}$ (black dotted
line) is the same (within error bars) for $H\|j$ and $H$$\perp j$.
Indeed, as we will see (Fig.\ \ref{Fig8} below), at $5\,{\rm K}$
{\em after cooling in zero field} the resistivity is only very
weakly field dependent for an in-plane field (both for transverse
and longitudinal configurations).

\begin{figure}[tb]
\includegraphics[width=0.98\linewidth]{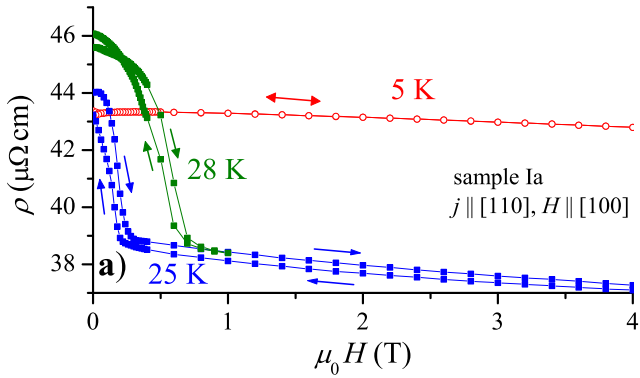}
\caption{(Color online) Resistivity vs $H\|j\|$[110] at selected
$T$ (sample Ia).} \label{Fig8}
\end{figure}

This indicates that the difference in the $\rho (T)$ curves of
Fig.\ \ref{Fig6} originates from the magnetism. The
magnetoresistance then may arise either from spin-disorder
scattering or scattering of domain walls\cite{Singh76} and the
difference can be due to a different configuration of domains. In
principle, this configuration (domain population, average size of
domains, etc.) may be different for $H\|[1\overline{1}0]$
(transverse magnetoresistance) and $H\|[110]$ (longitudinal
magnetoresistance) due to different demagnetizing effects (the
sample was cut in a rod such that the [110] direction is a bout
three times longer than the $[1\overline{1}0]$ direction). In any
case, the fact that the resistivity is almost field-independent
(for an in-plane field) at $5\,{\rm K}$ indicates that the
relevant (for resistivity) domain configuration in the LTP is very
hard to change by applying a magnetic field, whereas a field
applied in the ITP prepares a certain LTP domain configuration
upon cooling below $T_1$. We note that application of a field in
the ITP also seems to lead to small, but systematic, changes in
the resistivity values in the HTP, persisting up to $\sim$$T_N$.

Figure \ref{Fig7}a) shows the evolution of $\rho (T)$ [for clarity
only the data measured upon warming, after the lowest temperature
had been reached by cooling with the same field applied
(field-cooled-warming measurement protocol) are displayed] with
increasing $H\|j$ (again sample Ia, $j\|$[110]). It seems clear
that the $T_2$ transition systematically shifts to higher
temperatures with increasing $H$. The $T_1$ transition is far less
field dependent (for $H\|$[110]), and as discussed above, quite
hysteretic. Furthermore, the resistivity in the ITP becomes lower
in higher $H$, developing into a well-defined weakly
temperature-dependent valley above $\sim$$2\,{\rm T}$. This
development is also visible in the $\rho (H)$ dependence at
constant $T$ (Fig.\ \ref{Fig8}). There is only a minimal field
dependence of $\rho$ at $5\,{\rm K}$ in the LTP. In contrast to
this, a pronounced drop in low fields at $25\,{\rm K}$ in the ITP
is shifted to higher fields by increasing $T$ and seems to
transform into the drop associated with the ITP-HTP transition. In
the $\rho (T)$ curves of Fig.\ \ref{Fig7}a), for intermediate
field ranges, there are indications for an additional feature in
the ITP-HTP transition. The details of this feature have to be
clarified in future work.

The evolution of $\rho (T)$ with increasing $H\|$[001] for the
same sample and current density direction is shown in Fig.\
\ref{Fig7}b). The data shown were again obtained using the
field-cooled-warming measurement protocol, but we note that
application of $H\|$[001] (unlike $H\|$[110]) quickly suppresses
any hysteresis, and broadens the transitions. Apart from a
broadening, the $T_2$ transition does not seem to be strongly
affected. In contrast to this, $H\|$[001] systematically shifts
the $T_1$ transition to lower temperatures and above
$\sim$$1\,{\rm T}$ there are no longer any indications for a
transition to the LTP. This indicates that the LTP is suppressed
by $H\|$[001].

\subsection{Magnetization}
\label{highHM}

\begin{figure}[tb]
\includegraphics[width=0.98\linewidth]{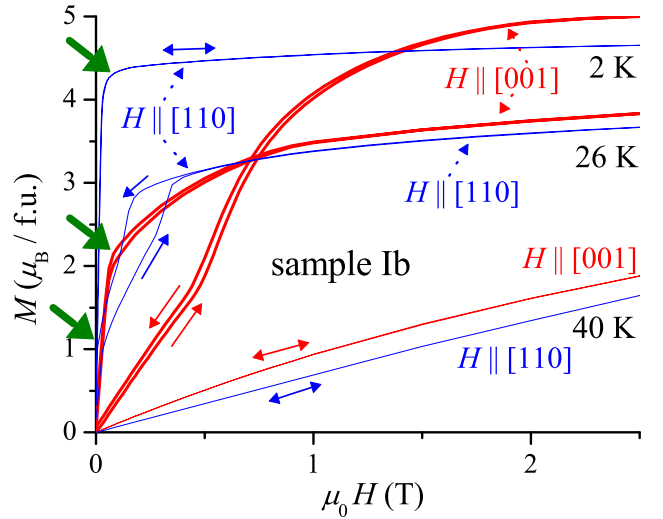}
\caption{(Color online) Magnetization $M$ as a function of field
$H$ at selected $T$, $H\|[110]$ (blue) and $H\|[001]$ (red),
sample Ib. Thin arrows indicate the $H$ direction of the
measurements, thick arrows indicate the criterion taken for the
estimation of the spontaneous ferromagnetic moment.} \label{Fig9}
\end{figure}

\begin{figure}[tb]
\includegraphics[width=0.98\linewidth]{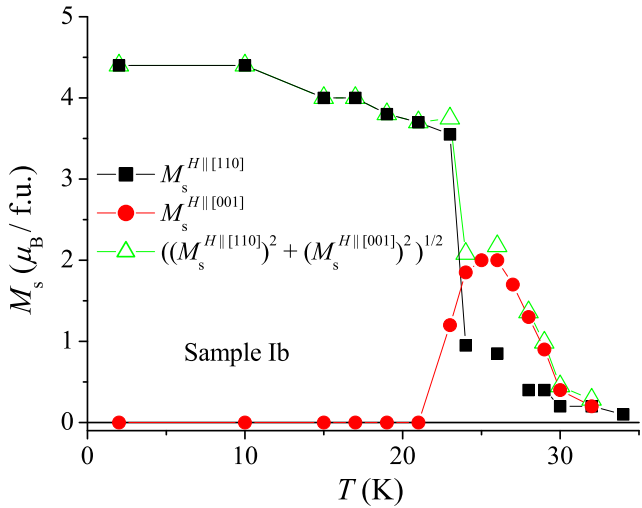}
\caption{(Color online) Spontaneous ferromagnetic moment $M_s$,
estimated from the $M(H)$ curves in Fig.\ \ref{Fig9} vs
temperature $T$, $H\|[110]$ (black squares) and $H\|[001]$ (red
circles), sample Ib. The square root of the sum of the squares for
both field directions is shown as open triangles.} \label{Fig9N}
\end{figure}

In order to investigate the magnetic nature of these phases, we
measured the magnetization $M$ as a function of $T$ and $H$
($\|$[110] and $\|$[001]). In Fig.\ \ref{Fig9}, we display for
sample Ib $M(H)$ loops at selected $T$. For $H\|$[110], at
$40\,{\rm K}$, $M(H)$ shows almost linear behavior, but at
$2\,{\rm K}$ $M(H)$ has a behavior typical for soft ferromagnets:
saturation in a field which is of the order of the estimated
demagnetizing field, but no hysteresis or remanence. However, the
saturation value of $4.5 \mu_B/{\rm f.u.}$ at $2\, {\rm K}$ is
much lower than the full Gd moment of $7 \mu_B/{\rm f.u.}$. At
$26\,{\rm K}$, pronounced hysteresis is visible in low $H$. This
is ubiquitous for $M(H)$ in the ITP and correlates with the
hysteresis in the drop in $\rho$ visible in Fig.\ \ref{Fig8}a).
The hysteresis may suggest a first-order metamagnetic transition.
In higher fields $M(H)$ is nearly constant, resembling the
behavior at $2\,{\rm K}$. In low fields $M(H)$ indicates a
spontaneous ferromagnetic moment, unlike at $2\,{\rm K}$, at
$26\,{\rm K}$ we did observe remanence. The $T$ dependence of the
spontaneous ferromagnetic moment, as estimated from Fig.\
\ref{Fig9} taking into account demagnetizing effects (we took the
magnetization at the first kink in low $H$, indicated by thick
arrows in Fig.\ \ref{Fig9}), is shown in Fig.\ \ref{Fig9N} (black
squares). The LTP-ITP transition in $M(H)$, $H\|$[110] is
accompanied by a drop of the spontaneous moment to $\sim \! \! 1\,
\mu_B$, and by the appearance of hysteresis.

For $H\|$[001] at $26\,{\rm K}$ (ITP), the form of $M(H)$ (Fig.\
\ref{Fig9} red curves) is consistent with a spontaneous
ferromagnetic moment of about $2\, \mu_B$. For the LTP consider
$M(H)$ at $2\,{\rm K}$. Below $0.5\, {\rm T}$, $M(H)$ is linear,
above it rises more strongly before slowly starting to saturate.
The field $H_1$ where $M(H)$ starts to increase faster than linear
is lower at higher $T$, extrapolating to zero at $T_1$. For
samples with a higher $T_1$, $H_1$ at low temperatures is
proportionally higher (not shown). Whereas the feature at $H_1$ as
defined above is rather sharp, the slow saturation of $M$ above
$H_1$ may be due to a small inhomogeneity of the Fe content in the
sample, as discussed in Sec.\ \ref{lowH}.

For $H\|$[001], the saturation value of $M \lesssim 5\, \mu_B/{\rm
f.u.}$ is higher than for $H\|$[110], but still far below the full
Gd $4f$ moment. Note that the difference to $7\, \mu_B/{\rm f.u.}$
is much too large to be explained by a negative contribution of
the Gd conduction electrons. The low saturated moment implies that
the magnetization does not correspond to the magnetization of a
simple ferromagnet. For $H\|$[001], the temperature dependence of
the estimated spontaneous ferromagnetic moment is shown in Fig.\
\ref{Fig9N} as red circles. It is similar to $M(T)$ in low $H$
(c.f.\ Fig.\ \ref{Fig4}).

The linear $M(H)$ behavior (below $0.5 \, {\rm T}$ at $2\,{\rm
K}$) may be due in-plane ferromagnetic moments resisting a
reorientation by an out-of-plane field. The LTP-ITP transition
could correspond to a reorientation of ferromagnetically aligned
$4f$ moments (as suggested in Ref.\ \onlinecite{Fujiwara87}).
Figure \ref{Fig9N} indicates that in the LTP the ferromagnetically
aligned moments are confined to the $a$$-$$b$ plane, whereas in
the ITP the moments are mainly out-of-plane, pointing into a
direction with an angle of about $25^{\circ}$ to [001]. A crude
estimate of the magnitude of the spontaneous moment is given by
$(M_{s,110}^2 + M_{s,001}^2)^{1/2}$, shown in Fig.\ \ref{Fig9N} as
open triangles. Apart from one point near the LTP-ITP transition
(where the estimation of $M_s$ is less clear than in other
regions) the temperature dependence of $(M_{s,110}^2 +
M_{s,001}^2)^{1/2}$ is consistent with the LTP-ITP transition not
affecting the magnitude of the ferromagnetic component, but only
it's direction. Finally, we note that, like the low-field
magnetization (Fig.\ \ref{Fig4} full line), the estimated
spontaneous moment extrapolates to zero around $30-35\,{\rm K}$,
i.e., slightly higher than $T_2$ as indicated by resistivity and
the specific heat maximum. This roughly correlates with the {\em
onset} of the $4f$ contribution to the specific heat (Fig.\
\ref{Fig5}), which is not sharp at all. We therefore think it is
not an indication of an additional transition, rather it may be
due to an inhomogeneous Fe distribution or due to significant
precursor-effects (e.g.\ fluctuations).

The electrical-transport and magnetization data presented in this
section will allow us to construct $H$$-$$T$ phase diagrams for
$H\|$[110] and $H\|$[001]. We will do this in Sec.\ \ref{disc}.
Furthermore, we have established that the Gd $4f$ moment order
setting in at $T_2$ has a large ferromagnetic component, and the
first-order (see Fig.\ \ref{Fig6}) transition at $T_1$ involves a
reorientation of the ferromagnetic component into the $a$$-$$b$
plane for $T<T_1$. In order to elucidate the underlying magnetic
structure we will now focus on scattering data.

\section{X-ray magnetic resonant scattering}
\label{xrms}

In order to prove the long-range nature of the magnetic order, and
to learn more about the magnetic structure than the information
that is obtainable from thermodynamic and transport measurements,
scattering techniques can be applied, typically neutron
scattering. For Gd148, neutron scattering is less feasible because
of the large neutron absorption cross-section of the natural
isotope of Gd. An alternative technique, which relies on intense
synchrotron radiation and exploits a large increase in the
scattered intensity due to resonant processes when the energy of
the radiation is close to an absorption edge of a magnetically
ordered element, is x-ray resonant magnetic scattering
(XRMS).\cite{Gibbs88,Hannon88,XRMSreview} To gain further insight
into the magnetism of Gd148, we therefore performed a first XRMS
experiment on the (together with piece Ia from the same crystal)
best characterized sample Ib.

\begin{figure}[t!b]
\includegraphics[width=0.98\linewidth]{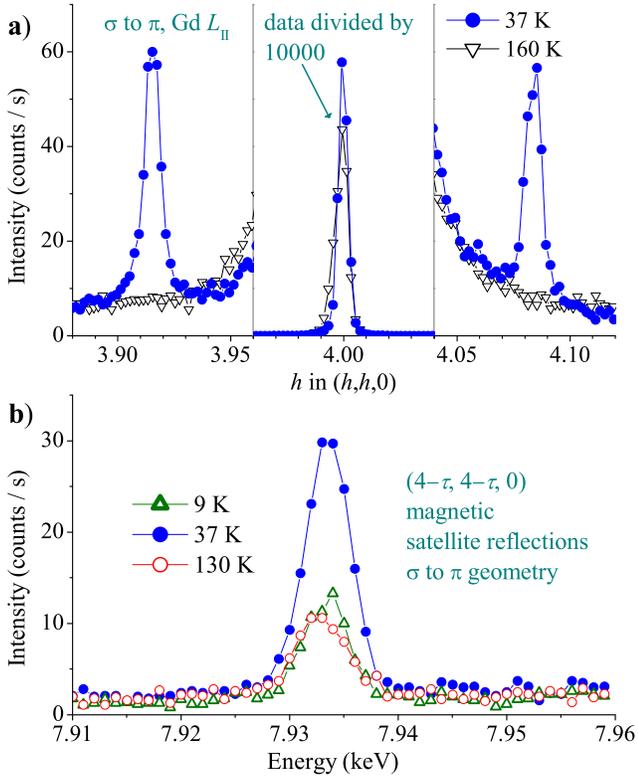}
\caption{(Color online) a) Reciprocal space scans along
[$h$,$h$,0] at $37$ and $160\,{\rm K}$ (sample Ib, Gd $L_{\rm II}$
edge). The data in the range $h=3.96$ to $4.04$ have been divided
by a factor of $10000$. %b) Rocking curves of the intensity ({\bf
%this Panel will be removed}). c)
b) Energy scans of the intensity of the ($4-\tau$,$4-\tau$,$0$)
satellite reflections at $9$, $37$, and $130\,{\rm K}$. Note that
intensities in panel a) are not comparable to intensities in panel
b).} \label{Fig11}
\end{figure}

Based on the neutron scattering and XRMS results on R148 with
other rare
earths,\cite{Langridge99,Paixao00,Paixao01,Schobinger98,Schobinger99}
as well as on the thermodynamic and transport results presented in
earlier sections, we expected at temperatures between $T_2$ and
$T_N$ a modulation of Fe moments with propagation along the [110]
direction to be present, with a feasible resonant enhancement of
the intensity scattered at the corresponding satellite reflections
at the Gd $L_{\rm II}$ edge. Figure \ref{Fig11}a) shows reciprocal
space scans along [$h$,$h$,$0$] with $\sigma \rightarrow \pi$
polarization analysis and the energy tuned to the Gd $L_{\rm II}$
edge ($7.934\,{\rm keV}$) at $37$ and $160\,{\rm K}$. At $37\,{\rm
K}$, the charge reflection at (4,4,0) is accompanied by small
satellites at ($4\pm \tau$,$4\pm \tau$,$0$) with $\tau=0.086$. The
satellite reflections are not present at $160\,{\rm K}$,
indicating that they are due to the magnetic ordering at the
{N{\'{e}}el} temperature (c.f.\ Fig.\ \ref{Fig2}). We also found
satellite reflections at ($3\pm \tau$,$3\pm \tau$,$0$), ($2\pm
\tau$,$2\pm \tau$,$0$), and ($1\pm \tau$,$1\pm \tau$,$0$), with
$0.06\leq \tau \leq 0.135$, depending on $T$. The full width at
half maximum (FWHM) of the satellite reflections is approximately
twice the FWHM of the charge reflections. The corresponding
smaller correlation length of the magnetic order may be an
indication of the presence of small domains. We roughly estimate
the average size of the domains (in [110] direction) to be
$\sim$$0.17\,\mu{\rm m}$ or about $140$ unit cells. Rearrangements
of these domains might be the cause of the small changes in
resistivity values in the HTP (see the discussion of Fig.\
\ref{Fig6} in Sec.\ \ref{highHrho}).

The intensity of the ($4\pm \tau$,$4\pm \tau$,$0$) satellite
reflections is about four orders of magnitude lower than the
intensity of the (4,4,0) charge reflection, despite the use of the
polarization analysis, which suppresses the charge reflection
intensity by three orders of magnitude. Due to the small intensity
of the satellite reflections and their vicinity to the charge
reflections, there is considerable intensity due to the charge
tail at the position of the satellite reflections. A slight
asymmetry in the charge tail leads to a larger charge contribution
to the scattered intensity at the ($4+\tau$,$4+\tau$,$0$)
reflection, and we therefore further analyzed only the
($4-\tau$,$4-\tau$,$0$) satellite reflection.

Figure \ref{Fig11}b) shows, at selected $T$, energy scans over the
Gd $L_{\rm II}$ edge for the satellite reflections at ($4-\tau ,
4-\tau , 0$). The intensity of the resonance is largest at
$37\,{\rm K}$. The lower intensity at $130\,{\rm K}$ is due to the
vicinity of the {N{\'{e}}el} transition, the lower intensity at
$9\,{\rm K}$ will be discussed below in Sec.\ \ref{disc}. The
position and the (amplitude-normalized) peak shape of the
resonance is $T$ independent within our resolution and consistent
with dipolar ($E1$)
transitions.\cite{Langridge99,Gibbs88,Hannon88,XRMSreview} We note
that a resonant enhancement at the Gd $L_{\rm II}$ edge was
observed only at the ($4\pm\tau , 4\pm\tau , 0$) satellite
reflections, there was no enhancement at the (4,4,0) charge
reflection, consistent with resonant magnetic scattering.
Reciprocal space scans and the resonant enhancement at the Gd
$L_{\rm II}$ edge are consistent with a magnetic modulation with
propagation vector ($\tau$,$\tau$,0) associated with Gd.

The specific-heat measurement discussed in Sec.\ \ref{lowH}
implies that in the HTP the Gd $4f$ moments are not ordered. Since
XRMS is element-sensitive, this might seem in contradiction to our
observation of a resonant magnetic signal at the Gd $L_{\rm II}$
edge at $37$ and $130\, {\rm K}$ (Fig.\ \ref{Fig11}). However,
XRMS ($E1$) of the Gd $L$ edges is sensitive to the polarization
of the Gd $5d$ bands, not (directly) to the $4f$
moments.\cite{Langridge99,Gibbs88,Hannon88,XRMSreview} A Gd $5d$
band polarization can also be induced by the Fe $3d$ moments. For
this situation, a very weak intensity of the corresponding
magnetic satellites is expected.\cite{Everitt95} Compared to other
Gd compounds,\cite{Granado04,Good05,Tan05,Kim05branch} the
intensity of the satellite reflections we observed is indeed very
weak $-$ more than three orders of magnitude lower. This low
intensity might be due to the ordered moments pointing into a
direction close to [001], since in our scattering geometry we are
not sensitive to moments in [001] direction. However, changing the
orientation of the sample such that [001] is in the scattering
plane\cite{Langridge99,Tan05} did not significantly change the
intensity, suggesting that its low value is not related to the
direction of magnetic moments. The behavior is similar to the
situation in Dy148 at high temperatures.\cite{Langridge99}

At this point it may be useful to briefly comment on an XRMS
signal coming directly from the Fe moments. This would of course
be very important to confirm the above considerations.
Unfortunately, the energies of the $L$ absorption edges of $3d$
transition metals do not allow diffraction experiments for regular
crystalline materials.\cite{Tonnerre95} Resonances at the $K$ edge
may be used instead, but the x-ray resonant process at transition
metal $K$ edges is neither well understood nor efficient. Reported
resonant enhancements at transition metal $K$ edges of magnetic
scattering intensities were typically very small and often broad
(see, e.g., Refs.\
\onlinecite{Hill97,Stunault99,Hill00,Neubeck01,Paolasini02}),
compared to the substantial enhancements often observed at rare
earth $L$ edges (see for example Refs.\
\onlinecite{Granado04,Good05,Tan05,Kim05branch}).

An additional problem in our case is the large charge scattering
background due to the vicinity of the magnetic satellite
reflections to the charge reflections. This is more problematic
for Fe $K$ edge measurements, because at the Fe $K$ edge, the
angle of the analyzer crystal for the polarization analysis is
further away from the ideal $90^{\circ}$ than in the case of the
Gd $L_{\rm II}$ edge, resulting in a significantly higher charge
scattering background at the Fe $K$ edge. Performing energy scans
in the range of the Fe $K$ edge ($7.11\,{\rm keV}$) we did not
observe a clear resonant enhancement. In view of the mentioned
difficulties, this does not, however, imply that the Fe $3d$
moments are not ordered. Rather, we conclude that the direct
observation of an XRMS signal from the Fe moments will need a
separate experiment with specialized equipment and enough time to
gather high statistics.

Although we did not observe a clear resonant enhancement at the Fe
$K$ edge, based on the above discussion of the intensity of the
XRMS at the Gd $L_{\rm II}$ edge and the comparison with other
R148 compounds, we conclude that the ($\tau ,\tau , 0$)
propagation vector describes the order of the Fe moments. A Fe
moment modulation along [110] is common for R148 compounds, but in
R148 with other rare earths $\tau$ is typically larger ($0.127 <
\tau < 0.243$) and varies less with
$T$.\cite{Paixao00,Paixao01,Schobinger98,Schobinger99}

At $9\,{\rm K}$, the satellite reflections are present as well
[Fig.\ \ref{Fig11}b)], with the propagation vector in the same
direction. This demonstrates that the Fe moments maintain their
modulation from the HTP, where the Gd $4f$ moments are not
ordered, in the LTP, where the Gd $4f$ moments {\em are} ordered.
This is similar to the situation in other R148 compounds. However,
the low intensity of the ($4-\tau , 4-\tau , 0$) satellite
reflections at low $T$ indicate that for Gd148, unlike the
situation in other R148
compounds,\cite{Langridge99,Paixao00,Schobinger98} the rare-earth
$4f$ moments do {\em not} follow the Fe moment modulation. In
order to find possible modulations involving the Gd $4f$ moments,
we searched at $9\,{\rm K}$ also for magnetic satellite
reflections in other high-symmetry directions. However, full scans
along [110], [100], and [001] (for scattering geometries mentioned
above) did not turn up any reflections with intensity high enough
to correspond to a modulation of a significant part of the $4f$
moments.

Further XRMS experiments, including tracking the observed magnetic
satellite reflections in detail through temperature and applied
field, and comparing the resonances at absorption edges other than
Gd $L_{\rm II}$ are planned.

\section{Discussion and conclusions}
\label{disc}

In the following we will discuss our results, starting with the
magnetic ordering at the {N{\'{e}}el} temperature and the nature
of the resulting phase. We will then discuss the $H-T$ phase
diagrams (for two field directions) at low temperature and the
magnetic order in these phases.

Our resistivity and magnetization data indicate that
GdFe$_4$Al$_8$ orders magnetically at $T_N \approx 155\,{\rm K}$
(see Sec.\ \ref{lowH}). Even small ($1-3\%$) iron deficiency
significantly lowers $T_N$, indicating that magnetic interactions
involving the Fe $3d$ moments drive the transition. In the x-ray
resonant magnetic scattering (XRMS) experiment (Sec.\ \ref{xrms}),
the order below $T_N$ was identified as an incommensurate
antiferromagnetic order with propagation in [110] direction.
Although corresponding satellite reflections were measured on the
Gd $L_{\rm II}$ edge, the $T$ dependence of entropy associated
with the Gd $4f$ moments, estimated from specific-heat data (see
Fig.\ \ref{Fig5} in Sec.\ \ref{lowH}), indicates that at high
temperature, the Gd $4f$ moments do not participate in this
antiferromagnetic order, which is thus driven by Fe-Fe moment
interactions. This can explain the weak intensity of the measured
satellite reflections. Antiferromagnetic order propagating in
[110] direction, at high $T$ confined to Fe $3d$ moments, seems to
be rather typical for $R$Fe$_4$Al$_8$ compounds.

In addition to the {N{\'{e}}el} transition, we found two more
phase transitions at low temperature ($T_1$ and $T_2$), dividing
the ordered state into three phases: LTP (low-temperature phase),
ITP (intermediate-temperature phase), and HTP (high-temperature
phase). Due to the proximity of the two low-$T$ transitions it
proved to be very useful to apply magnetic fields in two
directions and thereby more clearly separate these phases.

\begin{figure}[tb]
\includegraphics[width=0.98\linewidth]{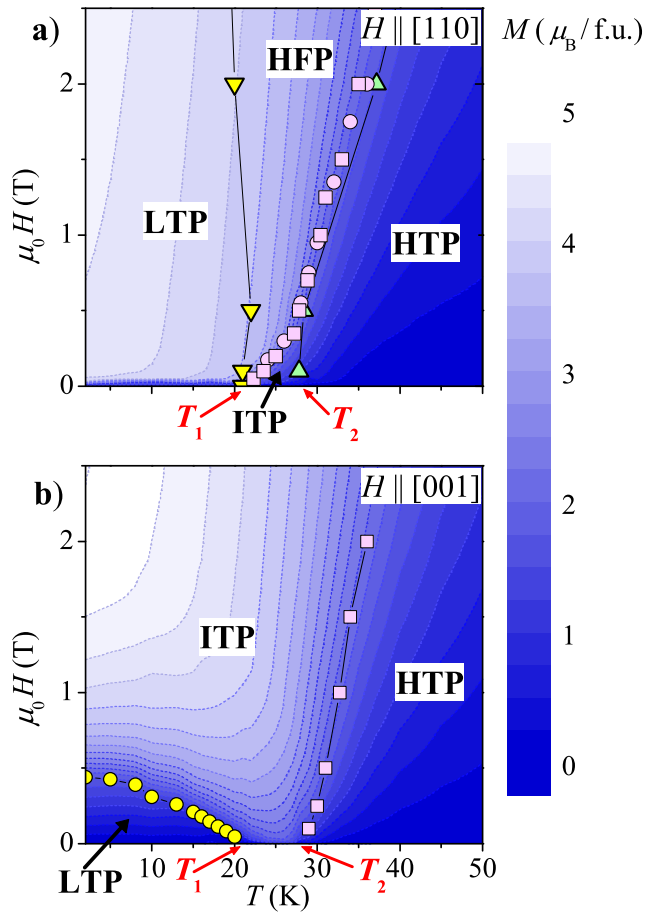}
\caption{(Color online) Low-temperature phase diagrams of
GdFe$_{3.96(1)}$Al$_8$ (sample Ia/b) for $H\|$[110] (a) and
$H\|$[001] (b). Triangles in a) are from resistivity measurements
on sample Ia, the other symbols from magnetization measurements on
sample Ib (see text for details). The three magnetic phases LTP
(low-temperature phase), ITP (intermediate-temperature phase), and
HTP (high-temperature phase), as well as the zero-field transition
temperatures $T_1$ and $T_2$ are indicated in both panels. A
possible fourth magnetic phase existing in fields applied parallel
to [110] (HFP) is indicated in panel a). Underlaid is a contour
plot of the magnetization $M$ (sample Ib).} \label{Fig10}
\end{figure}

Low-temperature phase diagrams established from resistivity and
and magnetization measurements (see Sec.\ \ref{highH}) on crystal
I (samples Ia and Ib) are presented in Fig.\ \ref{Fig10} [for
$H\|$[110] in panel a) and for $H\|$[001] in panel b)]. The phase
diagrams are underlaid with a contour-plot of the magnetization
[data determined from $M(H)$ measurements as displayed in Fig.\
\ref{Fig9}, for $H$ increasing].

For $H\|$[110] [Fig.\ \ref{Fig10}a)] the phase boundaries
($\bigtriangleup$,$\bigtriangledown$) were determined from $\rho
(T)$ measurements [steepest slope, c.f.\ Fig.\ \ref{Fig7}a); shown
is the average between points determined on curves measured with
$T$ increasing and decreasing]. In $M(T,H)$ the ITP-HTP transition
corresponds to, upon decreasing $T$, a steep rise in $M$ and
saturation. For definiteness, we used the extrema in ${\rm
d}M/{\rm d}H$ [${\rm d}(MT)/{\rm d}T$], plotted in Fig.\
\ref{Fig10}a) as $\circ$ ($\Box$). In case of hysteresis we again
took the average between $T$ or $H$ increasing and decreasing.
Above $0.5\,{\rm T}$, this tracks the ITP-HTP transition deduced
from $\rho (T)$ well, but in lower $H$ the smallness of the
magnetization component in [110] direction leads the transition
line established by these criteria two lower $T$, in fact
extrapolating to $T_1$ for $H=0$. The $M(H)$ curve measured at
$26\,{\rm K}$ (Fig.\ \ref{Fig9}) and the $\rho (H)$ curve measured
at $25\,{\rm K}$ (Fig.\ \ref{Fig8}) {\em do} suggest an additional
metamagnetic transition in the temperature region of the ITP. This
transition is what is determined from the criteria above below
$0.5\,{\rm T}$, starting from $T_1$ in zero field and apparently
merging with the boundary to the HTP at about $28\,{\rm K}$ in
$0.5\,{\rm T}$.

This seems to separate the ITP as deduced from resistivity
measurements into two phases, we call the tentative high field
phase HFP. In the ({\em longitudinal}, $H\|$[110]) magnetization
the HFP is indistinguishable from the LTP, and thus it could be
that the HFP is better described as a ``modified LTP'' phase
rather than as a ``modified ITP'' one. However, the electrical
transport data (see, e.g., Fig.\ \ref{Fig6}) clearly indicate a
first-order transition separating LTP and HFP. This transition
exhibits a wide hysteresis in temperature when $H\|$[110] is
substantial (see Fig.\ \ref{Fig6}) and likely involves complex
transformation of domains present in the LTP and in the HFP (see
the brief discussion in Sec.\ \ref{highHrho}; note that the HFP
was not yet introduced there) warranting further investigation.
The connection of the LTP-HFP transition with pronounced domain
effects might indicate that domains are a significant force
driving this transition, but presently this is tentative at best.
The large hysteresis associated with the LTP-HFP transition
complicates the determination of the equilibrium transition line.
Estimating the equilibrium transition by taking the average of the
transitions for $T$ increasing and decreasing [for $3.5\,{\rm T}$
this is indicated by an open circle in Fig.\ \ref{Fig6}, in the
phase diagram, the transition is indicated by open down triangles]
suggests that $H\|$[110] favors neither the LTP nor the HFP
strongly. Similarly, the transition between the ITP and the HTP is
hard to pin down from $M(T)$ measurements with $H\|$[110] because
of the smallness of the component along [110] of the
magnetization. We observed no second ``extremum in $[{\rm d}(MT) /
{\rm d}T]$'', using this criterion only leads to the transition
between the HFP and the ITP, rather than to the one between the
ITP and the HTP. For a field applied along [110], resistivity vs
temperature measurements clearly indicate two transitions, whereas
(longitudinal) magnetization vs temperature measurements indicate
only one. Only when examining the spontaneous magnetization along
[110] (Fig.\ \ref{Fig9N}) is the presence of two separate
transitions clearly indicated.

For $H\|$[001] [Fig.\ \ref{Fig10}b)], three distinct regions in
the contour-plot of the magnetization are readily identified with
LTP (low $M$), ITP (high $M$) and HTP (low $M$). The transition
lines between these phases are qualitatively consistent with what
may be deduced from the resistivity measurements presented in
Fig.\ \ref{Fig7}b). In contrast to the phase diagram for
$H\|$[110], there are no indications for the presence of
additional phases for $H\|$[001]. Since $H\|$[001] significantly
broadens the features in $\rho (T)$, we rely here on magnetization
measurements (see Figs.\ \ref{Fig4} and \ref{Fig9}) only to draw
the transition lines in Fig.\ \ref{Fig10}b). To fix the ITP-HTP
transition, we used the same procedure as for $H\|$[110]
(squares). For $H\|$[001] too, the $H$ dependence of the
transition and a saturation consistent (considering demagnetizing
effects) with a spontaneous moment indicate a ferromagnetic
component in the ITP (see also Fig.\ \ref{Fig9N}). The LTP-ITP
transition is determined from the $M(H)$ loops (c.f.\ Fig.\
\ref{Fig9}), taking the field value where $M$ starts to increase
faster than linear as the criterion for the transition field. This
transition is displayed by yellow circles in Fig.\ \ref{Fig10}b).
For $H\rightarrow 0$ the transition line extrapolates to $T_1$,
reaffirming it's identity as the LTP-ITP transition. It can be
seen that $H\|$[001] quickly suppresses the LTP.

Based on the specific-heat data, we can conclude that the HTP to
ITP transition corresponds to the ordering of the Gd $4f$ moments.
Since the magnetization data indicate the presence of a
spontaneous ferromagnetic moment in the ITP, this Gd $4f$ ordering
is either ferromagnetic or has a ferromagnetic component. The
increase of the transition temperature $T_2$ upon application of a
magnetic field (Fig.\ \ref{Fig10}) is consistent with this. Thus,
Gd moments do not simply follow the Fe moment order from higher
temperatures [this is in contrast to, e.g., DyFe$_4$Al$_8$ (Ref.\
\onlinecite{Paixao00})].

The transition from the LTP to the ITP seems to be of first-order
as indicated by the observation of thermal hysteresis in the
resistivity (see Fig.\ \ref{Fig3}) as well as the magnetization
(see Fig.\ \ref{Fig9}). Surprisingly though, no feature could be
identified in the specific heat corresponding to this transition
and any latent heat would have to be very small. The LTP-ITP
transition is associated with a lock-in of the observed
spontaneous ferromagnetic component (Fig.\ \ref{Fig9N}) into the
$a$$-$$b$ plane in the LTP, but there are no clear indications
that its magnitude is changed by the transition. The appearance of
an additional antiferromagnetic modulation is possible, at least
it is not clear why a pure lock-in transition of a ferromagnetic
component should result in a sizeable increase of the resistivity
(both for $j\|$[110] and $j\|$[001]).

The presence of the HFP in $H\|$[110] complicates the situation.
In fact, there are clear indications  (meta-stability) for the
first-order nature of the transitions at all boundaries of the
HFP. In contrast to this, there are no indications for
meta-stability in the ITP-HTP transition. Meta-stability seems to
be present in the LTP-ITP transition, but it is much weaker than
in any transition from or to the HFP. We point out that the
first-order transition lines LTP-HFP and HFP-HTP both seem to
extrapolate to $T_1$ in zero field, and therefore the observed
smaller meta-stability of the LTP-ITP transition might well be
connected with the HFP phase. This is a strong incentive for
further investigations of the HFP.

We now attempt to draw a picture accounting for the full magnetic
moment of both Gd and Fe. The full ordered moment of Gd is
$7\,{\mu_B}/{\rm Gd}$. Since the valence of Gd is fixed and there
are no crystal-electric-field effects, this is the same for
different compounds, with only small corrections, up to $\sim
0.5\,\mu_B$, due to contributions by itinerant electrons (such as
the Gd $5d$ electrons). The full moment of Fe can vary from
compound to compound.

We know from the specific heat and XRMS measurements that in the
HTP Gd moments (the Gd $4f$ moments at least) are not ordered, and
Fe moments are antiferromagnetically ordered with propagation
along [110]. Magnetization measurements further show that there is
no significant ferromagnetic component present [see the $40\,{\rm
K}$ $M(H)$ loops in Fig.\ \ref{Fig9}]. We recall that, according
to the x-ray-diffraction-structure refinement and specific-heat
measurements, the Gd lattice is fully occupied and that at low
temperatures the Gd $4f$ moments are well ordered. The appearance
of a spontaneous ferromagnetic moment below the temperature $T_2$
where the Gd moments start to order (compare Figs.\ \ref{Fig5} and
\ref{Fig9N}) indicates that the ordering of the Gd moments is
ferromagnetic, but the saturation moments for both field
directions are far below the $7\,{\mu_B}/{\rm f.u.}$ (see Fig.\
\ref{Fig9}) expected for a full ferromagnetic alignment of Gd
moments. What could be the reason for this?

One possibility is that the Gd $4f$ moments are not fully
ferromagnetically ordered, but rather also have an
antiferromagnetic component. However, since we did not find any
satellite reflections with high enough intensity to account for
the necessary fraction of the full Gd $4f$ moment in any of the
high symmetry directions (see Sec.\ \ref{xrms}, note that the
intensity of the ($4-\tau,4-\tau,0$) reflection is much too low)
this scenario is not very likely.

From high-field measurements on finely ground free powder
particles of Gd148 (likely, at least roughly, corresponding to
$H\|$[001], ITP), Duong {\em et al.}\cite{Duong01} suggested for
$H\rightarrow 0$ a ferrimagnetic structure with Gd moments
parallel, Fe moments antiparallel to the field, with an
antiferromagnetic Fe moment order re-established only in high
fields. On the one hand, such a simple ferrimagnetic arrangement
of Gd and Fe moments is inconsistent with our observation in the
XRMS experiment of the presence of the ($4-\tau,4-\tau,0$)
satellite reflection characteristic for the antiferromagnetic Fe
modulation even at $9\,{\rm K}$ [Fig.\ \ref{Fig11}b)]. On the
other hand, a related idea would be consistent with both the XRMS
and the magnetization measurements: upon the ferromagnetic
ordering of the Gd $4f$ moments, the Fe moments {\em could} pick
up a ferromagnetic component, leading to a canted
antiferromagnetic structure. Such a canted magnetic structure with
a large ferromagnetic component may also explain why the intensity
of the ($4-\tau,4-\tau,0$) satellite reflection is about three
times smaller at $9\,{\rm K}$ (in the LTP) than at $37\,{\rm K}$
(in the HTP). Furthermore, it would also allow Fe to influence the
low-$T$ transitions, as indicated by the influence of Fe
deficiencies on all characteristic temperatures (see Sec.\
\ref{lowH}). The results of a recent dichroism experiment (on a
crystal ground to powder), which will be discussed in detail
elsewhere, are consistent with such a scenario as well.

This picture covers, at least for the LTP and the HTP, all the
data presented in the paper, but without more detailed scattering
data it remains speculative. Particularly concerning the ITP and
HFP, the absence of scattering data makes detailed conclusions
elusive.

Since the direction of the spontaneous ferromagnetic moment is
different for the ITP and the LTP, magnetocrystalline anisotropy
(MCA) is likely one of the forces driving this transition. In rare
earth compounds, MCA typically arises from crystal-electric-field
effects. As we mentioned in Sec.\ \ref{intro}, this source of
anisotropy is absent for Gd. Significant MCA that was nevertheless
observed for both elemental Gd and Gd compounds has been ascribed
to a combination of dipole-dipole interactions and (mainly)
anisotropy in the $5d$ bands.\cite{ColarietiTosti03,KimPhD} The
corresponding magnetic-anisotropy energies are much lower than the
magnetic-anisotropy energies resulting from crystal-electric-field
effects in other rare earths. A weak magnetic-anisotropy energy
might also, at least partly, explain why no latent heat could be
observed at $T_1$. In Gd148, anisotropy in the Fe $3d$ bands might
be an additional source of MCA.\cite{note_Lu148} In any case, the
different spontaneous moment direction in the ITP likely results
in a different structure of the Fe moments. The apparent shift of
$T_1$ upon changing the Fe stoichiometry may be taken as an
indication of the involvement of Fe moments in the LTP-ITP
transition. Before going any further in the discussion of the
driving forces (there is likely a delicate balance between
multiple energy scales) of the transitions between LTP, ITP, and
HFP, the magnetic structures of these phases need to be solved.
With this aim in mind, additional scattering experiments are
planned.

Regardless of the details of the magnetic structure at low
temperature, it is important to point out that i) on the one hand,
the Fe moments have at least a component antiferromagnetically
modulated in [110] direction ii) The Gd $4f$ moments, on the other
hand can {\em not} have a component ordered with the same
propagation as the Fe $3d$ moments, because then the intensity of
the reflections measured at $9\,{\rm K}$ would have to be orders
of magnitude higher (see Sec.\ \ref{xrms}). This is strikingly
different from the behavior for example in Dy148, where the rare
earth moments order at lower temperature, but then follow the Fe
moment modulation, and it implies a {\em co-existence of two
distinct orders} associated with Gd $4f$ and Fe $3d$ moments.
Whereas compounds in which one magnetic sublattice orders at
temperatures much lower than the other magnetic sublattice are not
very rare (indeed this was well known to be the case for R148
compounds in general), we are aware of only one example,
PrBa$_2$Cu$_3$O$_{6.92}$,\cite{Hill00} in which the magnetic
sublattice ordering at lower $T$ has no component with the same
modulation that the other sublattice has at all $T$.

The at low temperatures persisting antiferromagnetic modulation of
the Fe moments coexisting with an order of the Gd moments with a
large ferromagnetic component and without a component having the
Fe moment modulation is the central point of our paper. It may
indicate a modified relation between the strength of the Gd-Gd and
Gd-Fe moment interactions in Gd148 as compared to other R148
compounds. As we mentioned, apart from the stronger couplings
between $4f$ and conduction electrons in general, the most
peculiar property of Gd is the absence of crystal-electric-field
effects.

It seems remarkable that in the absence of the MCA due to crystal
electric field effects, the magnetic behavior in Gd148 is {\em
more} complex than in other R148. Likely, strong
crystal-electric-field effects play a central role in the
magnetism in R148 (except for Gd148). E.g., in Dy148,
crystal-electric-field effects press the $4f$ moments into the
crystallographic plane, making an order following the Fe-moment
modulation more probable, since the Fe moments are already
in-plane. Our results on Gd148, in contrast, indicate a complex
interplay of local $4f$ and itinerant $3d$ moments when
crystal-electric-field effects are absent.

\section{Summary}
\label{conc}

We presented an extensive set of data including magnetization,
electrical transport, specific heat, and x-ray resonant magnetic
scattering, measured on flux-grown single crystals of
GdFe$_4$Al$_8$. We found that two transitions at $T_1 \sim 21\,
{\rm K}$ and $T_2 \sim 27\, {\rm K}$ at low temperature divide the
ordered state below $T_N \sim 155\, {\rm K}$, where Fe moments
order antiferromagnetically with modulation along [110], into
three phases. The corresponding phase diagrams for fields applied
in [110] and [001] direction were presented.

Gd $4f$ moments order, mainly ferromagnetically, below $T_2$.
Above $T_1$ the ferromagnetic component of the moments points into
a direction close to [001], whereas below $T_1$ the ferromagnetic
component is locked into the $a$$-$$b$ plane. The Gd $4f$ moment
order is distinct from the Fe-moment order At low temperature $T$:
the Fe moments still have at least a component modulated along
[110], but the Gd $4f$ moments apparently do not have such a
component. We proposed a canted antiferromagnetic structure of the
Fe moments at low $T$, which can cover all the low-$T$ data
presented in the paper, but to some extent still is speculative.
Above  The transition at $T_1$ likely involves a delicate balance
of multiple energy scales associated with both Gd and Fe.

The complex magnetism in GdFe$_4$Al$_8$ as compared to
$R$Fe$_4$Al$_8$ compounds with other rare earths $R$ likely is
related to a modified ratio of coupling strengths and to the
absence of crystal-electric-field effects. In order to further
elucidate the structure of the magnetic phases of GdFe$_4$Al$_8$
additional scattering experiments are planned.

\begin{acknowledgments}
\label{ack} We thank S.~L. {Bud'ko}, P.~Ryan, and R.~W. McCallum
for useful discussions, P.~Ryan also for technical assistance
during the synchrotron experiment. Ames Laboratory is operated for
the U.S. Department of Energy by Iowa State University under
Contract No.\ W-7405-Eng-82. This work was supported by the
Director for Energy Research, Office of Basic Energy Sciences.
Synchrotron work was performed at the MuCAT sector at the Advanced
Photon Source supported by the U.S. DOE, BES, and OS under
Contract No.\ W-31-109-Eng-38.
\end{acknowledgments}

\newcommand{\noopsort}[1]{} \newcommand{\printfirst}[2]{#1}
  \newcommand{\singleletter}[1]{#1} \newcommand{\switchargs}[2]{#2#1}


\begin{thebibliography}{10}

\bibitem{Campbell72}
I.~A. Campbell, J. Phys. F: Metal Phys. {\bf 2},  L47  (1972).

\bibitem{Buschow78}
K.~H.~J. Buschow and A.~M. {van~der~Kraan}, J. Phys. F: Met. Phys.
{\bf 8},
  921  (1978).

\bibitem{Felner78}
I. Felner and I. Nowik, J. Phys. Chem. Solids {\bf 39},  951
(1978).

\bibitem{Fujiwara87}
H. Fujiwara, W.-L. Liu, H. Kadomatsu, and T. Tokunaga, J. Magn.
Magn. Mater.
  {\bf 70},  301  (1987).

\bibitem{Duong01}
N.~P. Duong, E. {Br{\"u}ck}, F.~R. {de~Boer}, and K.~H.~J.
Buschow, Physica B
  {\bf 294-295},  212  (2001).

\bibitem{Duong02}
N.~P. Duong, E. {Br{\"u}ck}, P.~E. Brommer, A. {de~Visser}, F.~R.
{de~Boer},
  and K.~H.~J. Buschow, Phys. Rev. B {\bf 65},  020408(R)  (2001).

\bibitem{Langridge99}
S. Langridge, J.~A. {Paix{\~a}o}, N. Bernhoeft, C. Vettier, G.~H.
Lander, D.
  Gibbs, S.~A. {S{\o}rensen}, A. Stunault, D. Wermeille, and E. Talik, Phys.
  Rev. Lett. {\bf 82},  2187  (1999).

\bibitem{Paixao00}
J.~A. {Paix{\~a}o}, M.~Ramos Silva, S.~A. {S{\o}rensen},
B.~Lebech, G.~H. Lander, P.~J. Brown,
  S. Langridge, E. Talik, and A.~P. {Gon{\c c}alves}, Phys. Rev. B {\bf 61},
  6176  (2000).

\bibitem{Paixao01}
J.~A. {Paix{\~a}o}, M.~R. Silva, J.~C. Waerenborgh, A.~P. {Gon{\c
c}alves},
  G.~H. Lander, P.~J. Brown, M. Godinho, and P. Burlet, Phys. Rev. B {\bf 63},
  054410  (2001).

\bibitem{Schobinger98}
P. Schobinger-Papamantellos, K.~H.~J. Buschow, and C. Ritter, J.
Magn. Magn.
  Mater. {\bf 186},  21  (1998).

\bibitem{Schobinger99}
P. Schobinger-Papamantellos, K.~H.~J. Buschow, I.~H. Hagmusa,
F.~R. {de~Boer},
  C. Ritter, and F. Fauth, J. Magn. Magn. Mater. {\bf 202},  410  (1999).

\bibitem{Sol1}
Z. Fisk and J.~P. Remeika,  in {\em Handbook on the Physics and
Chemistry of
  Rare Earths} (Elsevier, Amsterdam, 1989), Vol.~12.

\bibitem{Sol2}
P.~C. Canfield and Z. Fisk, Phil. Mag. B {\bf 65},  1117  (1992).

\bibitem{Sol3}
P.~C. Canfield and I.~R. Fisher, J. Crystal Growth {\bf 225},  155
(2001).

\bibitem{Waerenborgh00}
J.~C. Waerenborgh, P. Salamakha, O. Sologub, A.~P. {Gon{\c
c}alves}, C.
  Cardoso, S. {S{\'e}rio}, M. Godinho, and M. Almeida, Chem. Mater. {\bf 12},
  1743  (2000).

\bibitem{Janssen05}
Y. Janssen, M. Angst, K.~W. Dennis, R.~W. McCallum, and P.~C.
Canfield,
  cond-mat/0506587.

\bibitem{note_Palasyuk04}
Consistent with measurements on polycrystalline Gd148. See A.~M.
Palasyuk,
  B.~{Ya.} Kotur, E.~Bauer, H.~Michor, and G.~Hilscher, J. Alloys Compd. {\bf
  367}, 205 (2004).

\bibitem{note_Fujiwara}
Indications for a transition around $20\, {\rm K}$ (but not its
first order
  nature) were earlier found by Fujiwara {\em et al.}\cite{Fujiwara87} Their
  torque measurements indicate a change of the angle between the magnetic easy
  axis and the $ab$ plane from $\sim 30^{\circ}$ at high $T$ to $0^{\circ}$ at
  low $T$.

\bibitem{Mackintosh62}
A.~R. Mackintosh, Phys. Rev. Lett. {\bf 9},  90  (1962).

\bibitem{note_normalizedM}
Due to demagnetization effects, the magnetization at the peak
maximum can vary
  drastically from sample to sample. Normalization allows to better compare
  transition temperatures.

\bibitem{Hagmusa98}
I.~H. Hagmusa, E. {Br{\"u}ck}, F.~R. {de~Boer}, and K.~H.~J.
Buschow, J. Alloys
  Compd. {\bf 278},  80  (1998).

\bibitem{note_DuongGdFe2Ge2}
See, e.g., N.~P. Duong, E.~{Br{\"u}ck}, P.~E. Brommer, J.~C.~P.
Klaasse, F.~R.
  {de~Boer}, and K. H. J. Buschow, J. Magn.\ Magn.\ Mater.\ {\bf 242-245}, 813
  (2002).

\bibitem{Harris95}
J.~M. Harris, Y.~F. Yan, P. Matl, N.~P. Ong, P.~W. Anderson, T.
Kimura, and K.
  Kitazawa, Phys. Rev. Lett. {\bf 75},  1391  (1995).

\bibitem{Singh76}
R.~L. Singh, Phys. Rev. B {\bf 14},  4106  (1976).

\bibitem{Gibbs88}
D. Gibbs, D.~R. Harshman, E.~D. Isaacs, D.~B. McWhan, D. Mills,
and C. Vettier,
  Phys. Rev. Lett. {\bf 61},  1241  (1988).

\bibitem{Hannon88}
J.~P. Hannon, G.~T. Trammell, M. Blume, and D. Gibbs, Phys. Rev.
Lett. {\bf
  61},  1245  (1988).

\bibitem{XRMSreview}
For a recent review see articles in Synchrotron Radiation News,
Vol.\ 14, No.\
  5 (2001).

\bibitem{Everitt95}
B.~A. Everitt, M.~B. Salamon, B.~J. Park, C.~P. Flynn, T.
Thurston, and D.
  Gibbs, Phys. Rev. Lett. {\bf 75},  3182  (1995).

\bibitem{Granado04}
E. Granado, P.~G. Pagliuso, C. Giles, R. {Lora-Serrano}, F.
Yokaichiya, and
  J.~L. Sarrao, Phys. Rev. B {\bf 69},  144411  (2004).

\bibitem{Good05}
W. Good, J. Kim, A.~I. Goldman, D. Wermeille, P.~C. Canfield, C.
Cunningham, Z.
  Islam, J.~C. Lang, G. Srajer, and I.~R. Fisher, Phys. Rev. B {\bf 71},
  224427  (2005).

\bibitem{Tan05}
L. Tan, A. Kreyssig, J.~W. Kim, A.~I. Goldman, R.~J. McQueeney, D.
Wermeille,
  B. Sieve, T.~A. Lograsso, D.~L. Schlagel, S.~L. {Bud'ko}, V.~K. Pecharsky,
  and K.~A. {Gschneidner,~Jr.}, Phys. Rev. B {\bf 71},  214408  (2005).

\bibitem{Kim05branch}
J.~W. Kim, Y. Lee, D. Wermeille, B. Sieve, L. Tan, S.~L. {Bud'ko},
S. Law,
  P.~C. Canfield, B.~N. Harmon, and A.~I. Goldman, Phys. Rev. B {\bf 72},
  064403  (2005).

\bibitem{Tonnerre95}
J.~M. Tonnerre, L. {S{\`e}ve}, D. Raoux, G. {Soulli{\'e}}, B.
Rodmacq, and P.
  Wolfers, Phys. Rev. Lett. {\bf 75},  740  (1995).

\bibitem{Hill97}
J.~P. Hill, C.-C. Kao, and D.~F. McMorrow, Phys. Rev. B {\bf 55},
R8662
  (1997).

\bibitem{Stunault99}
A. Stunault, F. {de~Bergevin}, D. Wermeille, C. Vettier,
{Th.~Br{\"u}ckel}, N.
  Bernhoeft, G.~J. McIntyre, and J.~Y. Henry, Phys. Rev. B {\bf 60},  10170
  (1999).

\bibitem{Hill00}
J.~P. Hill, D.~F. McMorrow, A.~T. Boothroyd, A. Stunault, C.
Vettier, L.~E.
  Berman, M. v.~Zimmermann, and T. Wolf, Phys. Rev. B {\bf 61},  1251  (2000).

\bibitem{Neubeck01}
W. Neubeck, C. Vettier, F. de~Bergevin, F. Yakhou, D. Mannix, O.
Bengone, M.
  Alouani, and A. Barbier, Phys. Rev. B {\bf 63},  134430  (2001).

\bibitem{Paolasini02}
L. Paolasini, R. Caciuffo, A. Sollier, P. Ghigna, and M.
Altarelli, Phys. Rev.
  Lett. {\bf 88},  106403  (2002).

\bibitem{ColarietiTosti03}
M. {Colarieti-Tosti}, S.~I. Simak, R. Ahuja, L. {Nordstr{\"o}m},
O. Eriksson,
  D. {{\AA}berg}, S. Edvardsson, and M.~S.~S. Brooks, Phys. Rev. Lett. {\bf
  91},  157201  (2003).

\bibitem{KimPhD}
J.-W. Kim, Ph.D. thesis, Iowa State University, 2005.

\bibitem{note_Lu148}
We note that in preliminary measurements, we observed significant
anisotropy in
  the dc magnetic susceptibility in the ordered state of Lu148, which seems to
  indicate that substantial {MCA} can arise from Fe alone.

\end{thebibliography}
\end{document}